\documentclass[prd,preprintnumbers,amsmath,amssymb,superscriptaddress]{revtex4}  
\usepackage{graphicx}
\usepackage{epsfig}
\usepackage{rotating}
\usepackage{dcolumn}
\usepackage{bm}
\newcommand{\beq}{\begin{eqnarray}} 
\newcommand{\eeq}{\end{eqnarray}} 

\begin{document}

\preprint{LPT-ORSAY-12-57}

\title{Reproducing the Higgs boson data with vector-like quarks}

\author{N.~Bonne, G.~Moreau \\
{\it Laboratoire de Physique Th\'eorique, B\^at. 210, CNRS,
Universit\'e Paris-sud 11 \\  F-91405 Orsay Cedex, France}}

\begin{abstract} 
Vector-Like (VL) quarks arise in the main alternatives to the supersymmetric extensions of the Standard Model (SM). 
Given the experimental possibility of a $125$~GeV Higgs boson with rates significantly different from the SM expectations, 
it is motivating to study the effects of VL quarks on the Higgs boson cross sections and branching ratios. We perform a systematic 
search for the minimal field contents and gauge group representations of VL quarks able to significantly improve the fit of the 
measured Higgs rates, and simultaneously, to satisfy the direct constraints on VL quark masses as well as the electro-weak precision tests. 
In particular, large enhancements can be achieved in certain diphoton channels -- as pointed out by both the ATLAS and CMS Collaborations -- 
optimizing then the Higgs rate fit. This is a consequence of the introduction of VL quarks, with high electric charges of $8/3$ or $-7/3$, which are
exchanged in the Higgs-to-diphoton loop. Interestingly, the field contents and formal Higgs couplings obtained here are similar to those of scenarios 
in warped/composite frameworks arising from different motivations. The various exotic-charge quarks predicted, possibly below the TeV scale, 
might lead to a rich phenomenology soon at the LHC.
\end{abstract} 

\maketitle

\section{Introduction}

The main drawback of the Standard Model (SM) is probably the gauge hierarchy problem
induced by the divergent quantum corrections to the mass of the Higgs boson~\cite{Englert:1964et,Higgs:1964ia,Higgs:1964pj,Guralnik:1964eu}. 
Various alternatives to the supersymmetry, from the AdS/CFT correspondence~\cite{Maldacena:1997re} paradigm, address this problem:  
the so-called little Higgs models~\cite{ArkaniHamed:2001nc,ArkaniHamed:2002qx,ArkaniHamed:2002qy},
the composite Higgs~\cite{Contino:2003ve,Agashe:2004rs,Agashe:2005dk,Contino:2006qr,Burdman:2007sx}
and composite top~\cite{Hill:1991at} scenarios, the Gauge-Higgs unification mechanism (see {\it e.g.} Ref.~\cite{Carena:2006bn,Carena:2007ua})
or the warped extra-dimension setup proposed by L.Randall and R.Sundrum (RS)~\cite{Gogberashvili:1998vx,Randall:1999ee}
-- and its well-motivated version with SM fields in the bulk~\cite{Gherghetta:2000qt} allowing to generate the fermion 
mass hierarchy~\cite{Huber:2000ie,Huber:2001ug,Huber:2002gp,Huber:2003sf,Chang:2005ya,Moreau:2006np,Moreau:2005kz,Agashe:2004cp,Agashe:2004ay,Agashe:2006wa,Agashe:2006iy,delAguila:2008iz,Raidal:2008jk,Grossman:1999ra,Appelquist:2002ft,Gherghetta:2003he,Moreau:2004qe}.
\\ 
All these frameworks predict the existence of additional color-triplet states with vector-like gauge couplings; these Vector-Like (VL) quarks 
arise~\footnote{VL leptons can also appear; it must be noted however that those are typically heavier than the VL quarks within the warped/composite models.
There is also the possibility of extra chiral fermions, like a fourth generation which is now quite constrained by the data on Higgs physics.} as
Kaluza-Klein (KK) excitations in the higher-dimensional scenarios, as excited resonances of the bounded states constituting the SM
particles in composite models and essentially as partners of the top quark being promoted to a larger multiplet in the little Higgs context. 
The VL quarks are also prevalent as KK excitations of bulk fields in other higher-dimensional frameworks (as {\it e.g.} in Ref.~\cite{Cheng:1999bg}) 
or as new components embedded into the simplest ${\rm SU(5)}$ representations in gauge coupling unification theories~\cite{Kilic:2010fs}.  
Hence VL quarks are predicted in most of the alternatives to supersymmetry and their masses can reach values below the TeV scale    
as it could occur {\it e.g.} for the so-called custodians, KK states arising in the custodially protected warped extra-dimension 
models~\cite{Agashe:2003zs,Agashe:2006at,Djouadi:2006rk,Djouadi:2007eg,Bouchart:2008vp,Bouchart:2009vq}~\footnote{Custodians might
even appear in warped extra-dimensional frameworks of supersymmetric models~\cite{Bouchart:2011va}.}.

Now from the experimental side, all the recent data on Higgs boson searches performed at the Tevatron and Large Hadron Collider 
(LHC) constitute an important insight in the exploration of the Electro-Weak Symmetry Breaking (EWSB) sector. 
Hints in the data collected in 2011 by the ATLAS and CMS experiments, which were observed in various final states, could correspond to slight excesses 
of events induced by a Higgs boson with a mass of $m_h \simeq 125$~GeV as announced in Ref.~\cite{CERNcouncil}. 
New and updated LHC results have appeared at the Moriond 2012 conference;  
then both LHC experiments reported local $\sim 3\sigma$ excesses~\cite{ATLASweb,CMSweb}
but more analyzed event statistics is still needed to confirm that these excesses with respect to the background estimation are not statistical fluctuations. 
The CDF and D0 Collaborations have also detected an excess around $\sim 125$~GeV in the bottom quark search channel, $h \to \bar b b$,  
corresponding to a local significance of $2.7\sigma$~\cite{TEVNPH:2012ab}.
\\
Under this hypothesis of the existence of a $125$~GeV Higgs boson, deviations with respect to the SM are observed:  
both ATLAS and CMS report an increase of all the 
Higgs production rates in the diphoton channel ($h \to \gamma\gamma$) compared to the SM Higgs rate, being close to the $2\sigma$ 
level in some cases [see the precise data in the following detailed discussion]. 
The combined CDF and D0 analyses point towards a best-fit cross section, for $p\bar p \to h V \to \bar b b V$ [$V\equiv Z^0,W^\pm$ bosons],  
centered at $\sim 2$ times the SM cross 
section~\cite{TEVNPH:2012ab} also for $m_h=125$~GeV~\footnote{The uncertainty on the $m_h$ resolution
is about $10$~GeV and the best-fit cross section decreases (increases) below (above) $125$~GeV.}. 
Finally, the Higgs production rate in the channel $h \to W^+W^-$ is below the SM rate by more than $1\sigma$
(still at $125$~GeV) in the CDF+D0, ATLAS and CMS results. 
The other measurements are in a reasonable agreement with the SM expectations.

The investigation of the Higgs sector, in addition to shed light on the cornerstone of the SM, 
represents a window on physics beyond the SM. The Higgs properties may {\it e.g.} help to discriminate between 
various new theories that could manifest themselves indirectly at the TeV scale.
In this spirit, it is interesting to try to explain the above deviations of the Higgs rates from their SM predictions on the basis of corrections
to the Higgs couplings. This approach is particularly relevant in the contexts of the well motivated VL quarks, discussed above, 
which are able to modify the Higgs interactions through fermion mixing and new loop contributions.

In this paper, we adopt the generic approach of considering VL quarks as possible manifestations of several classes of theories beyond the SM,
without specifying those theories. 
We elaborate the three types of minimal VL quark models (and optimize their parameters) allowing to correct the $125$~GeV Higgs couplings so that  
{\it (i)} most of the Higgs rates are significantly closer to their central measured values than the SM expectations and in particular for the channels 
$h \to \gamma\gamma, \bar b b, W^+W^-$,
{\it (ii)} all the Higgs rates belong to the $1\sigma$ experimental regions and in turn improve the global fit of the SM,
{\it (iii)} the direct limits on the masses of involved VL quarks are respected and their contributions to the EW precision observables are acceptable. 
\\ 
If the presently observed deviations from the SM Higgs rates are indeed to be explained by some VL quarks then our results predict the existence 
of at least two $b'$ states ($=$~VL quarks with the same electric charge 
$Q_{e.m.}$ as the $b$) plus at least either a pair of $q_{8/3}$ (VL quark with $Q_{e.m.}=+8/3$) 
accompanied by a $q_{5/3}$ or a pair of $q_{-7/3}$ with a $q_{-4/3}$. $t'$ components (same $Q_{e.m.}$ as the top) may arise too. 
Some realistic mass intervals obtained are~\footnote{For comparison, the most severe lower limit on an extra-quark is at $611$~GeV
(assuming the branching ratio for the relevant decay channel at unity)~\cite{Chatrchyan:2012yea}.}: 
$m_{q^1_{8/3}}\approx 650 - 1100$~GeV, $m_{q^1_{5/3}}\approx 800 - 1000$~GeV, 
$m_{q^1_{-7/3}}\approx 650 - 1000$~GeV, $m_{q^1_{-4/3}}\approx 800 - 900$~GeV, 
$m_{b_2}\approx 840$~GeV and $m_{t_2}\approx 900 - 1010$~GeV~\footnote{In 
our notations, {\it e.g.} $q^1_{8/3}$ denotes the lightest mass
eigenstate and $b_2$ the second lightest eigenstate (the lightest one being the observed bottom quark: rigorously $b_1$ but sometimes
noted just $b$ as usual).}.
We do not claim those values to be the only allowed ranges as such a conclusion would rely on exhaustive and thus long explorations 
of the parameter spaces which are quite large in the considered scenarios. 
Let us mention that for certain gauge group representations, the $q_{8/3}$ or $q_{-7/3}$ quark can be stable. 
\\ 
Interestingly, one should also remark that the essential field content and
couplings addressing the Higgs rate `anomalies', in all the models constructed here, also hold 
in specific set-ups of the custodially protected RS scenario (or composite model). 
In that sense, the present Higgs data could be seen as an indication for 
the RS scenario and these specific custodial set-ups. Note that this kind of custodial set-up [containing principally two $b'$ and two $q_{8/3}$ or 
$q_{-7/3}$ custodians] has not been considered so far in the literature on RS (composite) model phenomenology~\cite{Agashe:2003zs,Agashe:2006at}
-- including the implications for Higgs searches 
at colliders~\cite{Djouadi:2007fm,Bouchart:2009vq,Falkowski:2007hz,Casagrande:2010si,Azatov:2011qy,Carena:2012fk} with 
LHC inputs~\cite{Goertz:2011hj} --
except in Ref.~\cite{Bouchart:2008vp} (\cite{DaRold:2010as} for a dual composite analysis) -- disconnected from Higgs searches -- where it arises from some
considerations on the EW Precision Tests (EWPT).
\\ 
The VL field configurations that we point out could be realized in the context of any theory where the effects on the Higgs sector with a different origin 
from the VL quarks are negligible, like {\it e.g.} in RS scenarios with custodians around the TeV scale and decoupling KK excitations of the gauge bosons 
much above $\sim 3$~TeV (which is generally the order of the bound from EWPT~\cite{Agashe:2003zs,delAguila:2003bh,Cabrer:2011qb}); then the minimal realistic 
VL quark configurations and the optimized Higgs rates would be mainly identical as those obtained in the present simplified framework.
The other possibility is to have complete theories underlying the SM where VL quarks appear, together with {\it e.g.} light extra gauge bosons or an extended Higgs
sector, so that several types of effects could affect the SM Higgs rates, in which case our results could be useful for theorists as a guide on VL quark influences.

At this stage, one should mention related works. 
In Ref.~\cite{Carmi:2012yp} (see also references therein), it is shown that specific scenarios addressing the naturalness problem -- via the
presence of top-partners such as $t'$ -- like the little Higgs, multi-Higgs and pseudo-Goldstone Higgs boson (possible pattern of composite Higgs models) 
scenarios, are constrained in a non-trivial way by recent Higgs rate estimations [even before the Tevatron data]. 
Constraints from the recent Higgs data have also been imposed explicitly on the Minimal Composite Higgs Models with fermions embedded in spinorial or
fundamental representations of ${\rm SO(5)}$~\cite{Ellis:2012rx,Azatov:2012bz,Azatov:2012rd}.
Effective approaches constraining generic Higgs operators, including possible deviations to the EW gauge boson couplings 
(but no effects from additional fermions not mixed with the SM ones), can be found in Ref.~\cite{Espinosa:2012ir} (see Ref.~\cite{Degrande:2012gr}
for effective interactions involving both the top quark and Higgs field). 
Another study at the effective coupling level shows that 
ad-hoc massive fermions in color ${\rm SU(3)_c}$ representations up to the ${\bf 27}$ and with $\vert Q_{e.m.} \vert \leq 2$ 
can explain the simultaneously observed inequalities (in the author notations): $gV<1$, $V\gamma>2$, 
$g\gamma>1$~\cite{Barger:2012hv} ($\bar b b$ data from Tevatron not included)~\footnote{These authors demonstrate the No-Go theorem  
that no color-triplet fermion with $\vert Q_{e.m.} \vert \leq 1$ can realize $V\gamma>2$ and $g\gamma>1$. We escape it by introducing several
fermions with $\vert Q_{e.m.} \vert \geq 2$.}.
Some of the recent LHC investigations on the Higgs scalar have also been considered in completely different theoretical contexts such as 
the Universal Extra Dimensions, where the constraints from the $\gamma\gamma$ and $W^+W^-$ channels leave only an 
allowed narrow window near $m_h=125$~GeV~\cite{Brooijmans:2012yi}.
Within the minimal supersymmetric extensions of the SM, the corrections to the $hW^+W^-$ vertex due to the extended Higgs sector 
constitute also the main source of corrections to the loop-induced $h\gamma\gamma$ coupling, a problematic
correlation preventing to obtain opposite corrections to the
$W^+W^-$ and $\gamma\gamma$ rates as wanted today~\cite{Giardino:2012ww} (and references therein).
Finally, the fourth generation~\cite{Ishiwata:2011hr}, the radion~\cite{Cheung:2011nv} or the dilaton~\cite{Barger:2012hv} 
have difficulties to interpret the rate enhancement observed recently at LHC, for $m_h=125$~GeV,  
in the Vector Boson Fusion (VBF) channel: $VV \to h \to \gamma\gamma$, while a fermiophobic Higgs boson could
explain it~\cite{Gabrielli:2012yz}. Increasing considerably this channel rate is also doable by introducing new states with large electric charges
that can be exchanged in the $h\gamma\gamma$-loop, in the spirit of the highly-charged quarks introduced here; 
this idea was realized in the different context of the
type II see-saw mechanism where doubly-charged Higgs scalars arise~\cite{Arhrib:2011vc}~\footnote{One could also mention related studies
about another type of VL quark effects on Higgs searches: new channels of Higgs production through VL quark decays~\cite{Azatov:2012rj}.}.

In Section~\ref{modelbuild}, we discuss the effects of different models of VL quarks on the Higgs boson observables. Then these  
models and their parameter space are confronted to the collider data in Section~\ref{numfits}. We conclude in Section~\ref{conclu}.

\section{Model building}
\label{modelbuild} 

\subsection{The Higgs boson data}
\label{hdata}

We first define all the Higgs rate observables which have been measured at the Tevatron and LHC assuming a $125$~GeV Higgs boson. 
Those are the following signal strength modifiers (given with the last references for their experimental value): 
$\mu_{h\gamma}=\sigma_{\rm h}B_{\rm h\to \gamma\gamma}/\sigma_{\rm h}^{\rm SM}B_{\rm h\to 
\gamma\gamma}^{\rm SM}$~\cite{TEVNPH:2012ab,CONF-2012-019,PAS-HIG-12-001},
$\mu_{hV}=\sigma_{\rm h}B_{\rm h\to VV}/\sigma_{\rm h}^{\rm SM}B_{\rm h\to VV}^{\rm SM}$~\cite{TEVNPH:2012ab,CONF-2012-019,PAS-HIG-12-008},
$\mu_{h\tau}=\sigma_{\rm h}B_{\rm h\to \tau\tau}/\sigma_{\rm h}^{\rm SM}B_{\rm h\to \tau\tau}^{\rm SM}$~\cite{CONF-2012-019,PAS-HIG-12-008},
$\mu_{Vb}=\sigma_{\rm hV}B_{\rm h\to bb}/\sigma_{\rm hV}^{\rm SM}B_{\rm h\to bb}^{\rm SM}$~\cite{TEVNPH:2012ab,CONF-2012-015,CONF-2012-019,PAS-HIG-12-008}, 
$\mu_{qW}=\sigma_{\rm hqq}B_{\rm h\to WW}/\sigma_{\rm hqq}^{\rm SM}B_{\rm h\to WW}^{\rm SM}$~\cite{PAS-HIG-12-008} and
$\mu_{q\gamma}=\sigma_{\rm hqq}B_{\rm h\to \gamma\gamma}/\sigma_{\rm hqq}^{\rm SM}B_{\rm h\to \gamma\gamma}^{\rm SM}$~\cite{PAS-HIG-12-001,Farina:2012ea}, 
where $B$ stands for the branching ratios, $\sigma_h$ for the total cross section of the Higgs production (dominated by the
gluon-gluon fusion mechanism $gg\to h$), $\sigma_{hV}$ for the cross section of the Higgs production in association with a $V$-boson 
and $\sigma_{hqq}$ for the VBF rate. One needs to introduce also the quantity~\cite{CONF-2012-13,Giardino:2012ww},
$$
\mu_{X\gamma}=
\frac{\sigma_{\rm hX}}{\sigma_{\rm hX}^{\rm SM}}
\frac{B_{\rm h\to \gamma\gamma}}{B_{\rm h\to \gamma\gamma}^{\rm SM}} 
=
\frac{0.3 \times \sigma_{\rm gg\to h}+\sigma_{\rm hqq}+\sigma_{\rm hZ}+\sigma_{\rm hW}}{0.3 \times \sigma_{\rm gg\to h}^{\rm SM}
+\sigma_{\rm hqq}^{\rm SM}+\sigma_{\rm hZ}^{\rm SM}+\sigma_{\rm hW}^{\rm SM}}
\frac{B_{\rm h\to \gamma\gamma}}{B_{\rm h\to \gamma\gamma}^{\rm SM}} ,
$$
where the factor $0.3$ has been estimated recently from simulating additional QCD jets~\cite{Giardino:2012ww} to account for the  
efficiency of events issued from the gluon-gluon fusion to pass the cuts for the selection of 
the $hV$ and $hqq$ productions (basically on the Higgs boson transverse momentum).
Similarly, to be rigorous there is a factor of about $0.033$ for the suppression of the gluon-gluon fusion events (containing jets at NLO) by the dijet-class tagging
in the $\mu_{q\gamma}$ measurement; the uncertainty on this factor is of $70\%$~\cite{Chatrchyan:2012tw} but it does not alter significantly anyway the theoretical 
estimation of $\mu_{q\gamma}$ in our framework (given the $\sigma_{\rm gg\to h}$ corrections and the absence of $\sigma_{hqq}$ modifications).

The experimental values for all these $\mu$'s observables are synthesized in the right-part of Fig.(\ref{Fig:Model.II}) (or equivalently of Fig.(\ref{Fig:Model.I})).
In order to summarize the various results on these figures, we have combined under the basic gaussian assumption the ATLAS and CMS data 
for every search channel where both experiments provide results -- except for the $h\to \bar b b$ channel where we find
it more instructive to discuss the data separately (see next paragraph). 
The combination has been done without including the correlation; this is correct for the statistical error and the
uncorrelated systematic errors while the correlated systematic ones, like the theoretical uncertainty, are expected to be subleading compared 
to all the others~\cite{Carmi:2012yp} (so the way those are combined should not be crucial).
\\
What comes out at a first glance on the experimental results of Fig.(\ref{Fig:Model.II}) is, in particular, the enhancement of  
the estimated rates for the three diphoton channels compared to their respective SM predictions.
One notices also the significant reduction of $\sigma_{\rm h}B_{\rm h\to WW}$ relatively to its SM expectation,
observed simultaneously by the Tevatron and LHC. Concerning the $h\to \bar b b$ channel, if one does not consider the
ATLAS best-fit value which is negative, an enhancement appears with respect to the SM especially
at Tevatron where the obtained accuracy is better. The other channels fall into the $1\sigma$ regions.

\subsection{The VL quark effects}
\label{VLeff}

In order to improve the fit to the Higgs data, one would first need to increase the $\bar b b$ channel in Fig.(\ref{Fig:Model.II}) [right-part] with respect to the SM.
For that purpose, one needs to enhance the $h\to \bar b b$ branching fraction since
the presence of VL quarks does not induce large tree-level corrections to the $hVV$ vertex nor to the initial
$V\bar q q$ coupling involved in the $hV$ production. To increase the $\Gamma_{\rm h\to bb}$ width at $m_h=125$~GeV, 
at least two VL $b'$-like states (say $b'$ and $b''$), mixing together and with the SM $b$ state, 
need to be introduced so that the absolute value of the bottom Yukawa coupling can be increased thanks to additional elements in 
the Yukawa coupling matrix arising in the $(b, b',b'')$ basis~\footnote{Even if obvious, it is interesting to notice that VL leptons would not allow
to increase $\Gamma_{\rm h\to bb}$ by tree-level effects.}. At least a Yukawa coupling mixing $b$ and $b'$
and another one inducing the $b'-b''$ mixing are necessary; since the Higgs field is in a ${\rm SU(2)_L}$ doublet, gauge invariance imposes
that $b'$ and $b''$ must be embedded in different ${\rm SU(2)_L}$ representations, as well as for the Left-handed
$b_L$ state (or the $b_R$) and $b'$. Restricting to two $b'$-like states and to multiplets smaller than a triplet to have the minimal field content set-up
(extended set-ups do not bring different kinds of effects), one must thus embed $b'$ in a singlet and $b''$ in a doublet.

Similarly, to increase theoretically the diphoton channels in the $hqq$ and $hX$ productions
-- as suggested by Fig.(\ref{Fig:Model.II}) -- we have to enhance $B_{\rm h\to \gamma\gamma}$. 
One cannot increase $\sigma_{\rm gg\to h}$, to favor $\sigma_{\rm hX}$, since $\sigma_{\rm h}B_{\rm h\to \gamma\gamma}$ is already
significantly increased (from the $B_{\rm h\to \gamma\gamma}$ enhancement required by the $\sigma_{\rm hqq}B_{\rm h\to \gamma\gamma}$
data) so that in view of the data $\sigma_{\rm gg\to h}$ must decrease for compensating this large $\sigma_{\rm h}B_{\rm h\to \gamma\gamma}$ variation -- as
it will occur in this paper. 
Because of this aspect, the large experimental values of $\sigma_{\rm hqq}B_{\rm h\to \gamma\gamma}$ as well as $\sigma_{\rm hX}B_{\rm h\to \gamma\gamma}$
and the above $\Gamma_{\rm h\to bb}$ behavior, 
the needed theoretical enhancement of $B_{\rm h\to \gamma\gamma}$ via $\Gamma_{\rm h\to \gamma\gamma}$ has to be important. 
\\
The possible increase of the loop-induced $h\gamma\gamma$ coupling~\cite{Djouadi:2005gi} from the top quark loop-contribution, through a $t-t'$ mixing increasing 
largely the top Yukawa coupling, is not possible as this SM top coupling is already close to its perturbativity bound and moreover the dominant 
(and opposite-sign) triangular loop-contribution to $h\gamma\gamma$ remains to be from the $W^\pm$-boson exchange. 
Suppressing the top Yukawa coupling is not a solution neither, for increasing largely the $h\gamma\gamma$ coupling. 
\\ 
The last promising way to increase the $h\gamma\gamma$ coupling is to introduce VL quarks with high
electric charges leading to new loop-contributions favored by the two photon couplings. The two first possible exotic charges 
obtainable by extending ${\rm SU(2)_L}$ multiplets of $b',t'$ are $Q_{e.m.}=-4/3$ and $Q_{e.m.}=5/3$, as induced by the relation $Y=Q_{e.m.}-I_{3L}$ 
($Y\equiv$~hypercharge, $I_{3L}\equiv {\rm SU(2)_L}$~isospin), and have in turn absolute values higher than the top quark one.
However, given the present direct bounds around $600$~GeV [see below] for the masses of such VL quarks $q_{-4/3}$, $q_{5/3}$
(basically decaying like $t'$, $b'$ respectively: $q_{-4/3}\to bW$, $q_{5/3}\to tW$), 
we have found that their loop-contributions to the $h\gamma\gamma$ coupling do not reach large enough amounts in regard to Higgs fit
improvements. Then one has to introduce the next-higher absolute charges, held by $q_{-7/3}$ and $q_{8/3}$, to increase again the
electromagnetic couplings of the $h\gamma\gamma$ loop. The Yukawa couplings and masses of these $q_{-7/3}$, $q_{8/3}$ must be such that their
loop-amplitude interferes constructively with the $W^\pm$-boson exchange to generate enhancements.   
The condition for, say, the $q_{8/3}$ to be exchanged in the $h\gamma\gamma$ loop is clearly that it must couple directly   
to the Higgs boson; this means that there should be at least two $q_{8/3}$ components, noted $q_{8/3}$ and $q'_{8/3}$, 
belonging to different gauge representations. To be minimal in terms of field content and without loss of generality, 
we restrict ourselves to two $q_{8/3}$ (or two $q_{-7/3}$) components embedded in
${\rm SU(2)_L}$ representations up to triplets -- including those reveals another type of model with respect to the $q_{8/3}/q_{-7/3}$ decay. 
Besides, we do not consider charges, $\vert Q_{e.m.} \vert > 8/3$, as those do not bring effects of different nature and 
are less usual charges (even if $Q_{e.m.} = -10/3$ and $11/3$ are considered {\it e.g.} in Ref.~\cite{DaRold:2010as}). Therefore, the only possibilities are to
embed either $q_{8/3}$ in a singlet and $q'_{8/3}$ in a doublet or $q_{8/3}$ in a doublet and $q'_{8/3}$ in a triplet
(or similarly for $q_{-7/3}$).

A nice feature about the presence of highly-charged quarks is to greatly increase $\Gamma_{\rm h\to \gamma\gamma}$ and only this width,
since the diphoton channels `suffer' from some of the largest experimental discrepancies with the SM.

As another good consequence of the field configurations selected above, $\sigma_{\rm h}B_{\rm h\to WW}$ will be significantly reduced as it seems 
indeed to be indicated by the data  
(see Fig.(\ref{Fig:Model.II})): $B_{\rm h\to WW}$ is reduced due to the $\Gamma_{\rm h\to bb}$ increase and $\sigma_{\rm h}$ due to the destructive interference
between the $q_{8/3}$ (or $q_{-7/3}$) loop and the top quark loop contributing to $\sigma_{\rm gg\to h}$~\cite{Djouadi:2005gi}. Of course, this
latter feature of destructive interference with the top contribution~\footnote{This feature is rendered possible, in contrast with fourth generation quark models, by the
vectorial nature of the $q_{8/3}/q_{-7/3}$ whose mass origin does not reside exclusively in the EWSB.} 
must be preserved in the presence of additional VL $t'$ or $q_{-4/3}$ for instance.

\subsection{The minimal models}
\label{minmodels}

The combined theoretical conditions (discussed in Section~\ref{VLeff}) for improving the fit of the Higgs data 
(presented in Section~\ref{hdata}) lead to the following exhaustive list of minimal models for the VL quarks.
A first class of models, denoted as Models of type I, is defined by the following four possibilities for the field content:
\begin{equation} 
(q_{8/3},q_{5/3})^t_{13/6}\ , \ (q'_{8/3})_{8/3} \ , \ (b')_{-1/3} \ , \ (t',b'')^t_{1/6} \ \mbox{or} \ (b'',q_{-4/3})^t_{-5/6} \ , 
\label{Eq:model.I.1}
\end{equation}
\begin{equation} 
(q_{-4/3},q_{-7/3})^t_{-11/6} \ , \ (q'_{-7/3})_{-7/3} \ , \ (b')_{-1/3} \ , \ (t',b'')^t_{1/6} \ \mbox{or} \ (b'',q'_{-4/3})^t_{-5/6} \ ,
\label{Eq:model.I.2}
\end{equation}
\begin{equation} 
(q_{8/3},q_{5/3},t')^t_{5/3} \ , \ (q'_{8/3},q'_{5/3})^t_{13/6} \ , \ (b')_{-1/3} \ , \ (t',b'')^t_{1/6} \ \mbox{or} \ (b'',q_{-4/3})^t_{-5/6} \ ,
\label{Eq:model.I.3}
\end{equation}
\begin{equation} 
(b',q_{-4/3},q_{-7/3})^t_{-4/3} \ , \ (q'_{-4/3},q'_{-7/3})^t_{-11/6} \ , \ (b'')_{-1/3} \ , \ (t',b''')^t_{1/6} \ ,
\label{Eq:model.I.4}
\end{equation}
where we have written the field components in their transposed ${\rm SU(2)_L}$ group representations together with the hypercharge as a global subscript.
The Models I.(\ref{Eq:model.I.2}) (i.e. defined by the field content of Eq.(\ref{Eq:model.I.2})) and I.(\ref{Eq:model.I.4}) are characterized 
by a stable $q^1_{-7/3}$ in the case where $q^1_{-7/3}$ is the lightest field of all its multiplet partners 
(namely $q^1_{-4/3}$ and $b_2$). Indeed, the only potential decay channel
in I.(\ref{Eq:model.I.2}), $q^1_{-7/3}\to q^1_{-4/3} W^-$, would then be kinematically closed. In I.(\ref{Eq:model.I.4}),  
the decay channel through a virtual intermediate state, $q^1_{-7/3}\to q^{1\star}_{-4/3} W^-\to b_1 W^-W^-$,  
would be forbidden by the absence of $b-b'$ mixing (recall that $b\equiv$~SM bottom quark).
In I.(\ref{Eq:model.I.2}) and I.(\ref{Eq:model.I.4}), open decay channels for the $q_{-7/3}$ partners could be, $q^1_{-4/3} \to q^1_{-7/3} W^+$ and 
$b_2 \to q^{1(\star)}_{-4/3} W^+\to q^1_{-7/3} W^+W^+$, where $q^1_{-7/3}$ would then appear as missing energy at colliders.
\\
The comparable case where $q^1_{-4/3}$ or $b_2$ is the lightest field among all its multiplet partners (leading to a stable $q^1_{-4/3}$ or $b_2$) 
does not occur in the parameter space we will consider.
A similar discussion hold for $q_{8/3}$ and its partners within the Models I.(\ref{Eq:model.I.1}) and I.(\ref{Eq:model.I.3}).

A second class of models, Models II, is defined by these two possible field contents:
\begin{equation} 
(q_{8/3},q_{5/3},t')^t_{5/3} \ , \ (q'_{8/3},q'_{5/3})^t_{13/6} \ ,\ (q''_{5/3},t'')^t_{7/6} \ , \ (b')_{-1/3} \ , \ (t''',b'')^t_{1/6} \ \mbox{or} \ (b'',q_{-4/3})^t_{-5/6} \ .
\label{Eq:model.II}
\end{equation}
These models are characterized by the dominant decay channel of the highest-charge component: $q^1_{8/3} \to q_{5/3}^{1(\star)} W^+\to t_1 W^+ W^+$
(kinematically open in realistic frameworks), as allowed by the $t-t'$ mixing for which $(q''_{5/3},t'')$ has been added. 
The other possible decay into the $t_2$ instead of the $t_1$ eigenstate is subleading due to the phase space suppression induced by the $t_2$ mass 
-- or the three-body nature of the $q_{5/3}^{1(\star)}$ decay if $t_2$ is virtual.

The last type of models, Models III, is defined by,
\begin{equation} 
(b',q_{-4/3},q_{-7/3})^t_{-4/3} \ , \ (q'_{-4/3},q'_{-7/3})^t_{-11/6} \ ,\ (b'',q''_{-4/3})^t_{-5/6} \ , \ (b''')_{-1/3} \ \mbox{and/or} \ (t',b'''')^t_{1/6} \ .
\label{Eq:model.III}
\end{equation}
Here the dominant decay of the highest-charge component is $q^1_{-7/3} \to q_{-4/3}^{1(\star)} W^-\to b_1 W^- W^-$
(kinematically open), as induced by the $b-b'$ mixing -- allowed by the presence of $(b'',q''_{-4/3})$~\footnote{Note that this
multiplet contains a $b''$ that can play the r\^ole of the usual $b'''$ singlet for the bottom Yukawa enhancement.}. 
The similar decay obtained by replacing $b_1$  with $b_2$ has a much smaller rate.

At this stage, it is interesting to realize a certain theoretical consistence: all the minimal models obtained here    
are similar to concrete warped extra-dimension~\cite{Bouchart:2008vp} and their dual composite Higgs~\cite{DaRold:2010as} scenarios (constructed to satisfy EWPT),  
in the sense that these concrete scenarios also possess the above crucial features allowing to improve the Higgs rate fit.
Indeed, the representations I, II, III in Ref.~\cite{Bouchart:2008vp} (with the extension of Eq.(16) therein) or B2 in Ref.~\cite{DaRold:2010as} 
contain two $q_{-7/3}$ custodians coupled via a Yukawa term  
as well as two $b'$ custodians mixed together and with the $b$ quark through Higgs interactions,  
reflecting thus perfectly the VL quark configuration of the present Models I.(\ref{Eq:model.I.2}), I.(\ref{Eq:model.I.4}), III. 
Furthermore, the embeddings IV of Ref.~\cite{Bouchart:2008vp} or T3 of Ref.~\cite{DaRold:2010as} have two $q_{8/3}$ custodians with a Yukawa coupling 
as well as two $b'$ custodians with the required mixings,
exactly as for the quark set-ups here in Models I.(\ref{Eq:model.I.1}), I.(\ref{Eq:model.I.3}), II. 
The additional fields and mixings arising in these concrete realizations (like $q_{-4/3}$, $q_{5/3}$ states or heavy KK towers) are not expected to perturb 
drastically the potential Higgs rate ameliorations, and on the contrary, could even add more freedom. 
Besides, considering here all in all a unique set of VL fields -- not a replica per generation -- corresponds to the assumption in Ref.~\cite{Bouchart:2008vp,DaRold:2010as} 
where typically the custodians for the first two quark (and three lepton) SM generations decouple.

We end up this subsection by writing explicitly the Lagrangian for one of these models.
With the field content in Eq.(\ref{Eq:model.II}) for the Model II with say the $(b'',q_{-4/3})^t_{-5/6}$ doublet, all the possible mass
terms and Yukawa couplings appearing in the generic Lagrangian, invariant under the ${\rm SU(3)_c \! \times \! SU(2)_L \! \times \! U(1)_Y} $ gauge symmetry, are,
\begin{equation} 
{\cal L}_{\rm II} \ =  
\ Y \overline{\left ( \begin{array}{c}
t \\  b 
\end{array} \right )}_{L}
H^{\dagger} t^c_{R} + 
Y' \overline{\left ( \begin{array}{c}
q''_{5/3} \\  t'' 
\end{array} \right )}_{L}
H t^c_{R} +
Y_{8/3} \overline{\left ( \begin{array}{c}
q'_{8/3} \\  q'_{5/3} 
\end{array} \right )}_{L/R}
H \left ( \begin{array}{c}
q_{8/3} \\  q_{5/3} \\t'
\end{array} \right )_{R/L} \! +
Y_{5/3} \overline{\left ( \begin{array}{c}
q''_{5/3} \\  t'' 
\end{array} \right )}_{L/R}
H^{\dagger} \left ( \begin{array}{c}
q_{8/3} \\  q_{5/3} \\t'
\end{array} \right )_{R/L} 
\nonumber
\end{equation}
\begin{equation} 
+ \ Y_b \overline{\left ( \begin{array}{c}
t \\  b 
\end{array} \right )}_{L}
H b^c_R +
Y'_b  \overline{\left ( \begin{array}{c}
t\\  b 
\end{array} \right )}_{L}
H b'_{R}+
Y''_b  \overline{\left ( \begin{array}{c}
b''\\  q_{-4/3} 
\end{array} \right )}_{L}
H^{\dagger} b^c _{R}+
Y_{-1/3}  \overline{\left ( \begin{array}{c}
b''\\  q_{-4/3} 
\end{array} \right )}_{L/R}
H^{\dagger} b' _{R/L}
+ m \ \bar b'_L b^c_R+ m' \ \bar b'_{L} b'_{R} 
\nonumber
\end{equation}
\begin{equation} 
+\ m_{-4/3} \overline{\left ( \begin{array}{c}
b'' \\  q_{-4/3} 
\end{array} \right )}_{L} \! \!
\left ( \begin{array}{c}
b'' \\  q_{-4/3} 
\end{array} \right )_{R}
\! +\  m_{5/3} \overline{\left ( \begin{array}{c}
q''_{5/3} \\  t'' 
\end{array} \right )}_{L} \! \!
\left ( \begin{array}{c}
q''_{5/3} \\  t'' 
\end{array} \right )_{R}
\! +\ m'_{8/3} \overline{\left ( \begin{array}{c}
q'_{8/3} \\  q'_{5/3} 
\end{array} \right )}_{L} \! \!
\left ( \begin{array}{c}
q'_{8/3} \\  q'_{5/3} 
\end{array} \right )_{R}
\! +\ m_{8/3} \overline{\left ( \begin{array}{c}
q_{8/3} \\  q_{5/3}\\t' 
\end{array} \right )}_{L} \! \!
\left ( \begin{array}{c}
q_{8/3} \\  q_{5/3}\\t' 
\end{array} \right )_{R} \!
+ {\rm H.c.}
\label{Eq:LagModII}
\end{equation}
where $H$ represents the SM Higgs doublet, $L/R$ the fermion chiralities, the $Y$'s dimensionless Yukawa coupling constants
and the $m$'s various VL quark masses. Let us remark that the $Y_{8/3}$, $Y_{5/3}$ and $Y_{-1/3}$ terms could each be split into two terms
with different chirality configurations and coupling constants.
A field redefinition rotating $b^c_R$ and $b'_R$ allows to eliminate the $m$ term without loss of generality.
The Yukawa couplings for the first two up-quark generations are not written in the Lagrangian~(\ref{Eq:LagModII}) as their mixings with the top-partners $t',t''$
should be much smaller than the $t$-$t'$,$t''$ mixing. Indeed, new heavy $t'$-like states are closer in mass to the top quark and the top
is in general more intimately connected to the ultraviolet physics, like in warped/composite frameworks. 
Since the CKM mixing angles~\cite{Nakamura:2010zzi} are typically small, the first two up-quark flavors
should essentially decouple from the sector $t,t',t''$. A similar discussion hold including the down-quark sector
and the $b'$, $b''$ components~\footnote{The $t'$ or $b'$ states could 
contribute to Flavor Changing Neutral Current (FCNC) reactions which are experimentally well constrained; theoretically   
these FCNC contributions rely precisely on the whole SM set of Yukawa coupling constants for quarks. 
The treatment of such an high degree of freedom in the parameter space is beyond the scope of our study.}.

\section{Fitting the Higgs boson rates}
\label{numfits}

\subsection{The theoretical parameter space}
\label{parspace}

\begin{figure}[t]
\begin{center}
\begin{tabular}{cc}
\includegraphics[width=0.4\textwidth,height=6.5cm]{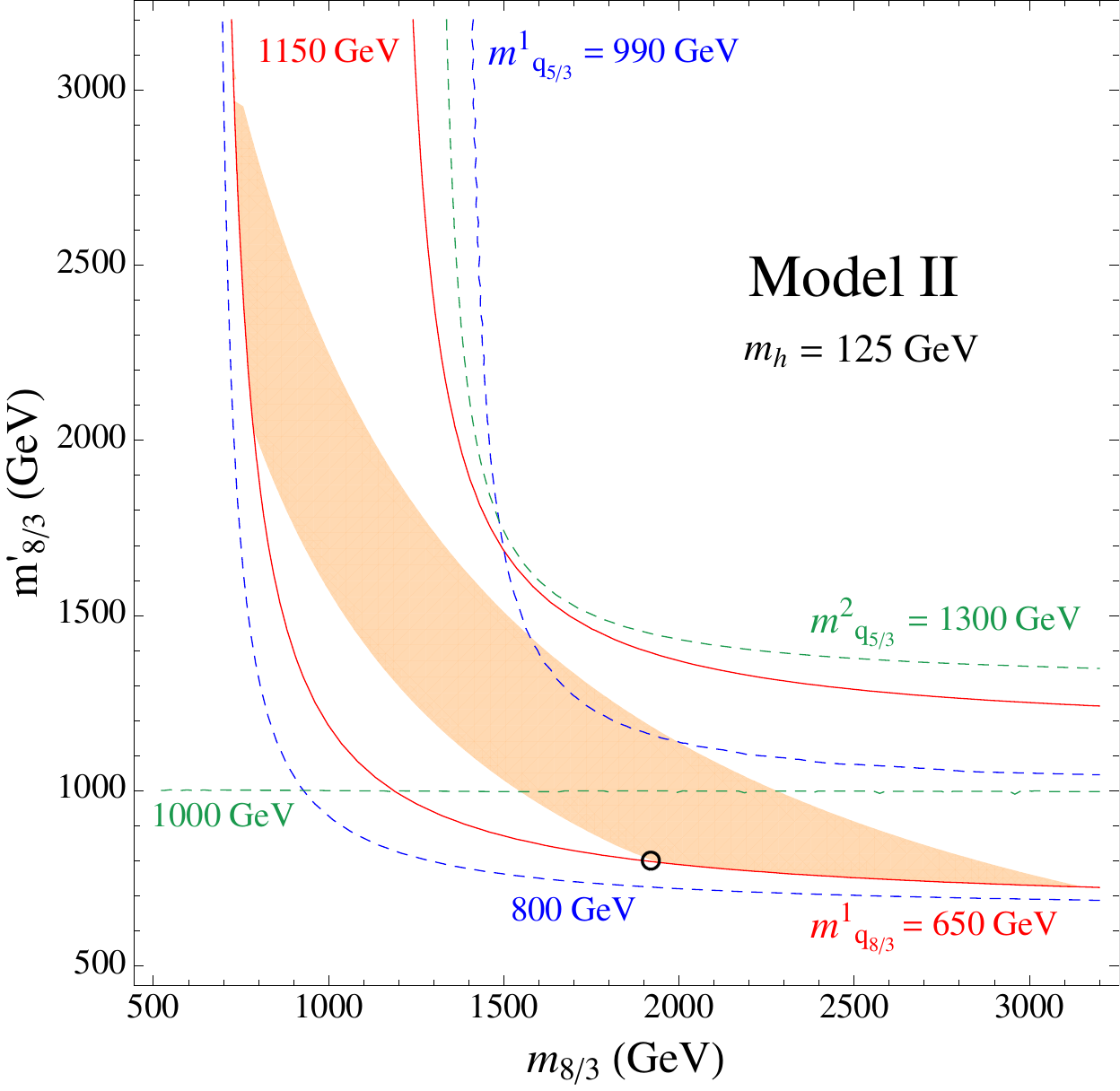}
\hspace{0.5cm}
&
\includegraphics[width=0.55\textwidth,height=6.5cm]{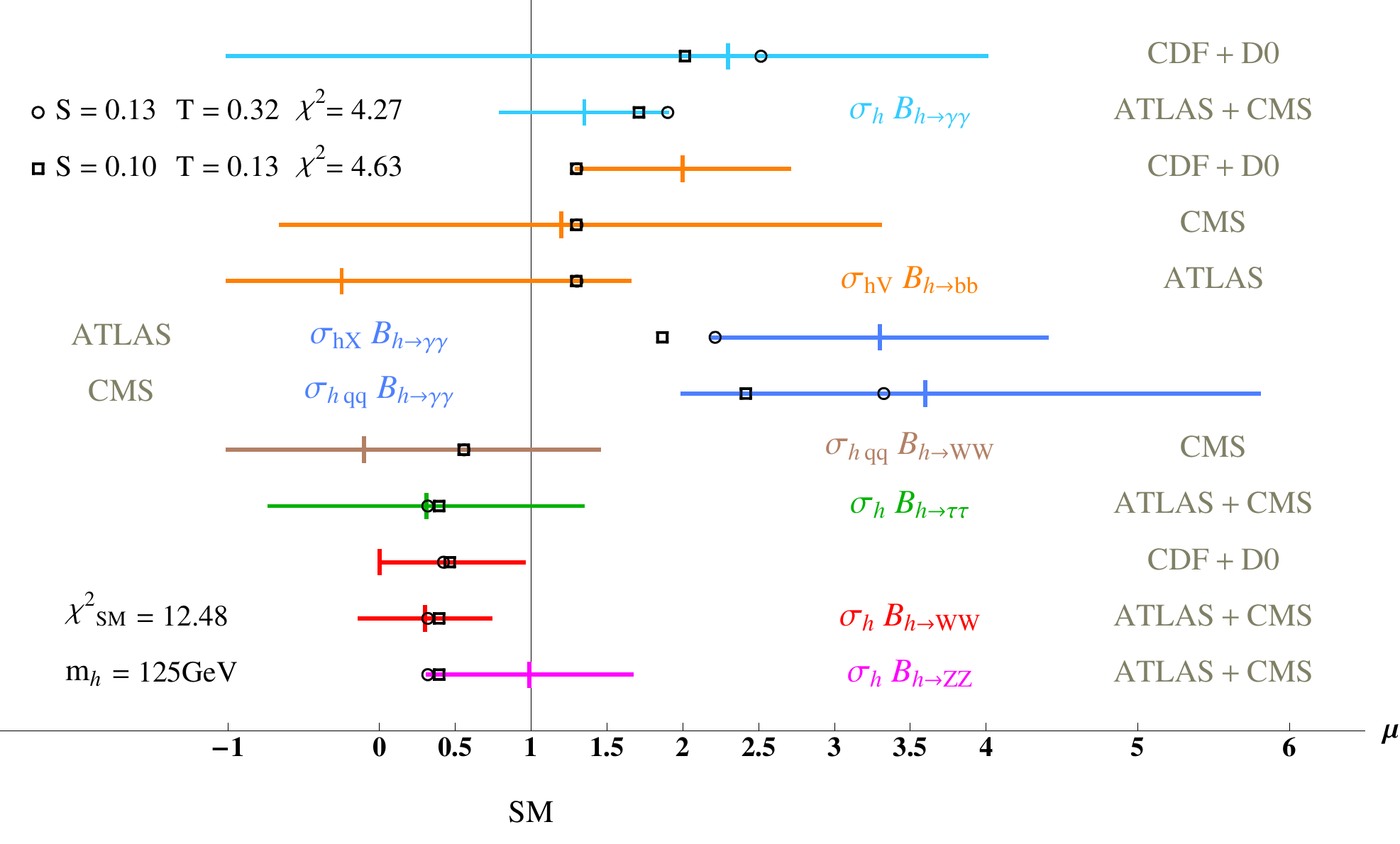}
\\
\end{tabular}
\caption{
\underline{Left:}  
Domain leading to $125$~GeV~Higgs boson rates inside the experimental $1\sigma$ intervals 
for the Model II, with the $(b'',q_{-4/3})$ doublet ({\it c.f.} Eq.(\ref{Eq:LagModII})), in the plan $m_{8/3}$ versus $m'_{8/3}$ (in GeV). 
The values of the other parameters are fixed at $Y=1.01$, $Y'=1$, $Y_{8/3}=2.5$, $Y_{5/3}=-0.5$,  $Y_b =-0.053$, $Y'_b =1$, $Y''_b=1$, 
$Y_{-1/3}=1$, $m'=1200$~GeV, $m_{-4/3}=900$~GeV, $m_{5/3}=1000$~GeV.
Contour-level curves for the physical masses $m_{q^1_{8/3}}$ and $m_{q^{1,2}_{5/3}}$ are also shown.
The other mass eigenvalues are almost constant over the shown plan like, $m_{b_1}\approx 4$~GeV, $m_{b_2}\approx 840$~GeV, 
$m_{b_3}\approx 1290$~GeV, $m_{t_1}\approx 173$~GeV, $m_{q^1_{-4/3}}\approx 900$~GeV, 
or mainly depending on $m_{8/3}$ like, $m_{t_2}\approx 900 - 1010$~GeV in the domain shown,  
or around $2$~TeV in this domain: $m_{t_3}\approx 1250 - 3000$~GeV, $m_{q^3_{5/3}}\approx 1500 - 3000$~GeV, $m_{q^2_{8/3}}\approx 1600 - 3000$~GeV.
\underline{Right:}  
Central values and $1\sigma$ error bars for the strength modifiers $\mu_{h\gamma}$, $\mu_{hV}$, $\mu_{h\tau}$, $\mu_{Vb}$, $\mu_{qW}$, $\mu_{q\gamma}$ 
and $\mu_{X\gamma}$ (defined in Section~\ref{hdata}) measured by the experiments indicated in front, for $m_h=125$~GeV. 
The various strength modifiers are indicated by the associated cross sections and branching ratios. The plus symbols mean that
the experimental results are combined. In each case the SM prediction corresponds to $\mu = 1$ leading to the global
$\chi_{\rm SM}^2$ value written in the figure. The small black circles correspond to the theoretical predictions of the strength modifiers for the point of parameter space
also indicated as a circle on the left-side plot. This parameter set leads to the oblique $S,T$ and Higgs fit $\chi^2$ values indicated 
near the circle on the top-left part of the figure. The little black squares are associated to a second parameter set where only one of the parameter values is changed:  
$Y_{8/3}=2.2$.}
\label{Fig:Model.II}
\end{center}
\end{figure}

We first consider the Model II which is quite attractive. 
In the left-part of Fig.(\ref{Fig:Model.II}), we present a domain of the parameter space where all the theoretical values of the Higgs rates belong to the experimental
$1\sigma$ regions [in the sense of Section~\ref{hdata}] which are shown on the right-part of Fig.(\ref{Fig:Model.II}). 
The Model II considered in this figure contains the $(b'',q_{-4/3})$ doublet of Eq.(\ref{Eq:model.II})
and its fundamental parameters appear in Eq.(\ref{Eq:LagModII}).
For this parameter space exploration, we have typically let the relative $\mu_{X\gamma}$ rate lying within a still acceptable 
$1.4\sigma$ region, to take into account an uncertainty in the QCD simulation of the efficiency for the gluon-gluon fusion contribution 
({\it c.f.} Section~\ref{hdata}). As the theoretical $\mu$'s quantities are normalized to the SM prediction, the QCD corrections arising in the VL quark contribution 
should essentially compensate the QCD corrections of the SM rate. 
\\ 
Within the domain of parameter space presented in the left-part of Fig.(\ref{Fig:Model.II}), we observe that all the VL quark masses are well above their 
strongest direct experimental constraints which are at most, 
$m_{b_2}>611$~GeV (with the conservative assumption $B_{b_2\to t_1W}=1$)~\cite{Chatrchyan:2012yea},  
$m_{t_2}>560$~GeV (again with $B_{t_2\to b_1W}=1$)~\cite{CMS:2012ab},  
$m_{q^1_{5/3}}>611$~GeV (assuming $B_{q^1_{5/3}\to t_1W}\simeq1$, like a $b_2$ state, which is a good approximation
as the channel $q^1_{5/3}\to t_2^{(\star)}W$ is subleading)~\cite{Chatrchyan:2012yea} and
$m_{q^1_{-4/3}}>560$~GeV (assuming similarly $B_{q^1_{-4/3}\to b_1W}\simeq1$, as a $t_2$ state, in a good approximation)~\cite{CMS:2012ab}.
There are no existing searches so far for a $q^1_{8/3}$ particle with the uncommon main decay, $q^1_{8/3} \to t_1 W^+ W^+$
(see discussion after Eq.(\ref{Eq:model.II})); anyway its mass values are quite high as illustrates Fig.(\ref{Fig:Model.II}) -- 
we have taken $650$~GeV as the lower limit -- and even higher for $q^2_{8/3}$.
The $q^2_{5/3}$ mass eigenstate decays 
either like the $q^1_{5/3}$ or as $q^2_{5/3}\to q^1_{5/3} Z$, $q^1_{5/3} h$ leading to a final state which has not been searched so far.
Identical considerations hold for the $q^3_{5/3}$, $b_3$ and $t_3$ eigenstates.
\\
Concerning the couplings, 
all the absolute values of the fundamental Yukawa parameters entering Eq.(\ref{Eq:LagModII}) have been taken larger than $0.5$ not to introduce new unexplained hierarchies
with respect to the top Yukawa coupling, $Y$, whose amount is close to unity as in the SM. The fundamental input parameter that is
the bottom Yukawa coupling, $Y_b$, has also the same order of magnitude~\footnote{A precise reproduction of the bottom and top quark masses
would require to include mixings with the first two generations.} 
as in the SM (even if the absolute physical coupling is slightly enhanced typically by $b-b'$ mixings to increase the $\bar b b$ decay channel).
The absolute Yukawa couplings in the mass basis do not exceed $2.5$ and are thus below the usual perturbativity upper bound at $\sqrt{4\pi}$.

The acceptable domain in the left-part of Fig.(\ref{Fig:Model.II}) is typically bounded from below by the $1\sigma$ constraint on $\mu_{hZ}$ 
and from above by the condition on $\mu_{X\gamma}$. This behavior of the Higgs boson rates is essentially due to the decoupling limit
where $m_{8/3}$ and $m'_{8/3}$ tend to high values in which the rates tend to their SM predictions. Similarly, on the figure, 
the independence of the smallest mass eigenvalue $m_{q^1_{8/3}}$ from $m_{8/3}$, at high values of the latter relatively to $m'_{8/3}$,  
is explained by the decoupling effect of $m_{8/3}$ in the mass matrix [and reciprocally for $m'_{8/3}$]. It is also the case for $m_{q^1_{5/3}}$ but
not exactly for $m_{q^2_{5/3}}$ as this mass matrix also involves $m_{5/3}$. 
\\ 
It is remarkable that the domain in Fig.(\ref{Fig:Model.II}), leading to Higgs rates in a good agreement with the present data, 
is relatively large. Similar domains arise for different values of the parameters which have been fixed for drawing this figure.

Let us finish this subsection by discussing the indirect constraints on the VL quarks. 
For the third generation quark sector, the tree-level corrections induced by the $t-t'$ ($b-b'$) mixings  
on the $t$ ($b$) vertex are expected to dominate over the loop-level oblique corrections to the
gauge boson propagators. 
Because of the relative heaviness of $t'$ states, the predicted value for the $V_{tb}$ CKM matrix element, including the $t-t'$ mixings, 
agrees with the experimental measurement obtained (without assuming $3\times 3$ unitarity) through the single top
production study~\cite{Nakamura:2010zzi}. In relation with the $Z\bar b b$ vertex, one could also try to address the LEP anomaly on
the Forward-Backward asymmetry for the bottom quark as done in the specific RS context~\cite{Djouadi:2006rk,Djouadi:2007eg,Djouadi:2009nb,Djouadi:2011aj}, 
assuming a discrepancy not due to under-estimated experimental errors, but this is beyond our scope.
\\Ê
Concerning the interactions of leptons and first generations of quarks, one has to compute the corrections to the gauge boson vacuum
polarizations induced at one loop by exchanged VL quarks~\cite{Barbieri:2006bg,Lavoura:1992np} in the present model. The values of the oblique parameters 
$S,T$~\footnote{$S,T$ encode the new physics effects only so that those vanish for the pure SM case.}
that we can reach belong to the $1\sigma$ regions induced by the long list of EW precision 
observables~\footnote{Remind that three crucial types of EW observables restricting the plan $S$ versus $T$ are $m_W$, $\Gamma_{Z\to\ell\ell}$ 
and the asymmetries ($\sin^2\theta_W$).} measured mainly at LEP~\cite{Nakamura:2010zzi}.
This is true in particular for the parameters inside the domain of Fig.(\ref{Fig:Model.II}) typically 
down to $m_{8/3}\simeq 1900$~GeV. Moving down to $m_{8/3}\sim 1000$~GeV along this domain  
leads to $T$ values up to $\sim 0.5$ (with $S\sim 0.1-0.2$) which would need to be compensated
by other new physics effects than the VL quark ones. Such effects might be induced by new fields heavier than the VL quarks so that the former would
correct $T$ (which is extremely sensitive to new physics effects via EW observables measured typically at the per mille level) but would leave the quality of the improved 
fit to Higgs rates mainly unaffected (given its present large error bars typically of several ten's of percents). 
Such a scenario could be realized {\it e.g.} in a warped framework with relatively light custodians, heavy KK fermionic towers,
heavy KK gauge bosons and possibly a gauge custodial symmetry (protecting the $T$ parameter) in the bulk.

\subsection{The fits versus oblique parameters}
\label{versus}

In the right-part of Fig.(\ref{Fig:Model.II}), we show the theoretical predictions of the Higgs boson rates still within the Model II
for a point of parameter space first optimizing
the fit -- for a same fit quality then choosing the smallest $S,T$ -- (black circles) and for another point favoring the oblique parameters $S,T$ (black squares). 
Both points are compatible with the direct constraints on VL quark masses.
\\ 
The circle-points in this figure show that it is possible to obtain Higgs rates being all within the $1\sigma$ regions. In this case, the fit is optimized in the following
sense: starting from this situation where the four predictions for $\mu_{h\gamma}$, $\mu_{hZ}$, $\mu_{Vb}$ and $\mu_{X\gamma}$ are at the extreme $1\sigma$ distances,
one cannot improve one of these four quantities without moving another one of those out of its $1\sigma$ region.   
One way of seeing this is as follows; the only possibility to increase $\mu_{X\gamma}$ while keeping $\mu_{h\gamma}$ at $1\sigma$ is to increase
$B_{\rm h\to \gamma\gamma}$ and decrease $\sigma_{\rm h}$. Now this $\sigma_{\rm h}$ decrease would worsen the fit on $\mu_{hZ}$ as the only
possible compensation by a $B_{\rm h\to ZZ}$ increase via a $\Gamma_{\rm h\to bb}$ decrease is forbidden if $\mu_{Vb}$ is to stay at the 
$1\sigma$ level.
\\ 
The conclusion of this feature is that other parameters reaching the same quality of Higgs rate fit as above can be found but there exist no parameters improving the fit
by comparison with the optimized situation described in the previous paragraph. Furthermore, 
neither different/additional ${\rm SU(2)_L}$ multiplets nor higher electric charges of VL quarks -- relatively to the present minimal model --
could improve the fit in that sense~\footnote{Of course {\it e.g.}
increasing the electric charge of VL quarks could allow to access larger VL quark masses achieving identical goodness of fits.}.
These conclusions are the consequences of a certain tension among the Higgs rate data which restricts a little bit the potential fit ameliorations 
brought by VL quarks (whatever are the model and parameters).

Let us also mention here that improving the Higgs fit through an enhancement of $B_{\rm h\to \gamma\gamma}$ with a simultaneous
suppression of $\sigma_{\rm h}$ -- as induced here by VL quark effects -- 
seems to be favored by the generic analysis at the level of cross sections and branching fractions~\cite{Giardino:2012ww}.
From a theoretical point of view, improving the fit on $\mu_{q\gamma}$ by
increasing the $hVV$ coupling -- from other origins of effects than the VL quarks -- is at the price of extending the Higgs sector~\cite{Falkowski:2012vh}. 
However, modifying this vertex, possibly through a custodial symmetry breaking~\cite{Farina:2012ea}, is probably the only way to reproduce the low $\mu_{hW}$
experimental value without affecting too much the perfect agreement on $\mu_{hZ}$ of the SM [{\it c.f.} Fig.(\ref{Fig:Model.II})].

The square-points in Fig.(\ref{Fig:Model.II}) correspond to $S,T$ values clearly inside the ellipse associated to the $1\sigma$ domain for combined EWPT~\cite{Nakamura:2010zzi}.
These points improve the fits of all the Higgs rates compared to the SM, especially $\mu_{Vb}$, $\mu_{X\gamma}$, $\mu_{q\gamma}$, $\mu_{hW}$ and with the exception of 
$\mu_{h\gamma}$ (same typical deviation from data as in the SM), $\mu_{hZ}$ (larger but still acceptable deviation). 
This configuration leads to a clear improvement of the global $\chi^2$ function compared to the SM, as written in the figure [for $\mu_{h\gamma}$, $\mu_{hW}$ only the most precise
value, i.e. from ATLAS+CMS, is included in $\chi^2$ whereas for $\mu_{Vb}$ only CDF+D0]~\footnote{Correlations are neglected.}.
In addition to improvements similar to those, the circle-points all belong to the $1\sigma$ regions of the Higgs rates with a remarkable increase for $\mu_{q\gamma}$; 
nevertheless, the associated $S,T$ values are slightly outside the $1\sigma$ 
ellipse~\cite{Nakamura:2010zzi}. There might however be other kinds of 
effects from the physics underlying the SM responsible for such necessary small $T$ compensations ($S$ value being not problematic here)
which could even let essentially unchanged the Higgs fit induced by the VL quark effects -- as described at the end of Section~\ref{parspace}.
Therefore, a noticeable but acceptable tension appears between optimizing the Higgs rate fit by introducing highly-charged VL quarks and still respecting the EWPT.

\subsection{The other models}
\label{othermod}

\begin{figure}[t]
\begin{center}
\begin{tabular}{cc}
\includegraphics[width=0.4\textwidth,height=6.5cm]{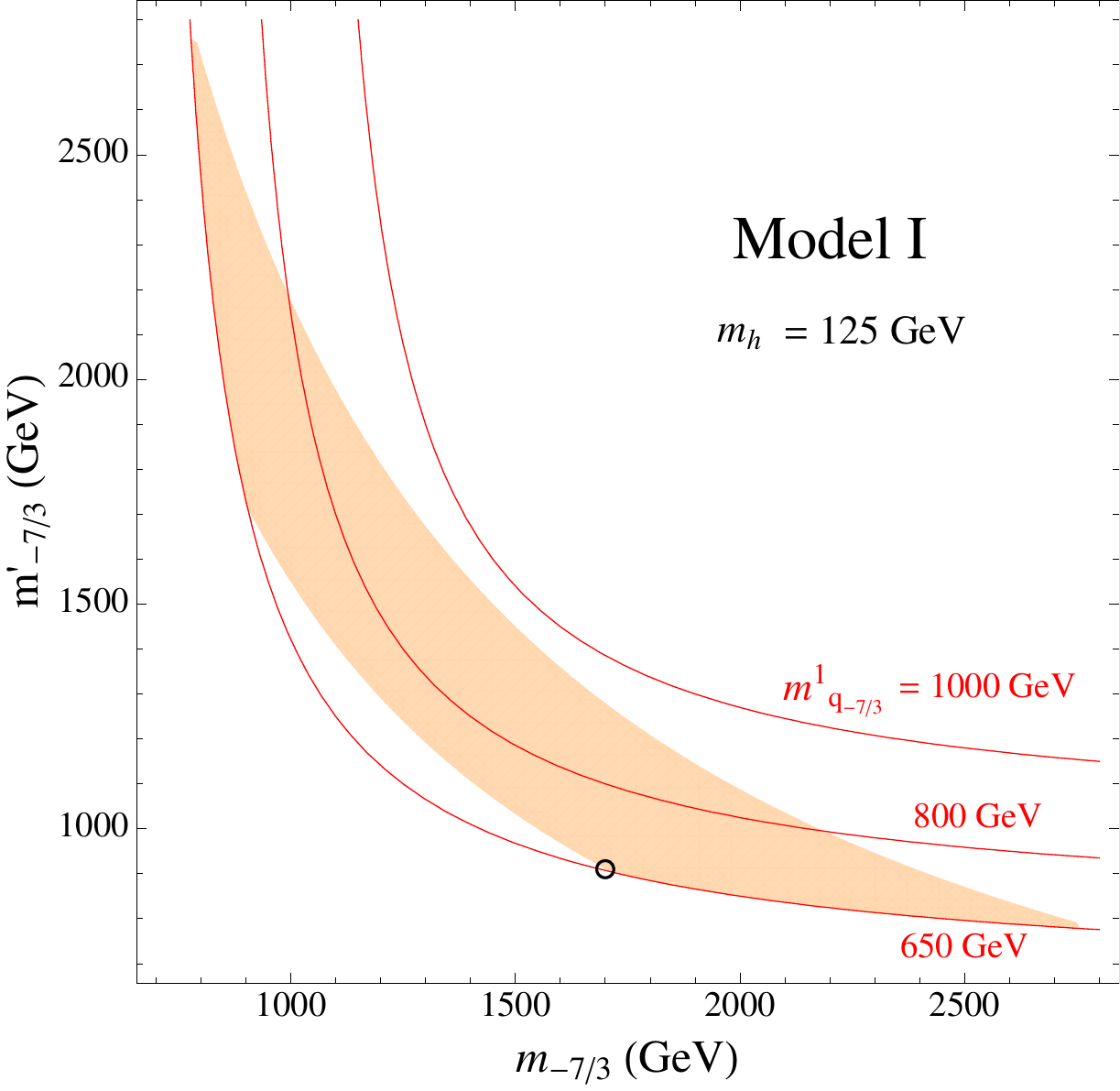}
\hspace{0.5cm}
&
\includegraphics[width=0.55\textwidth,height=6.5cm]{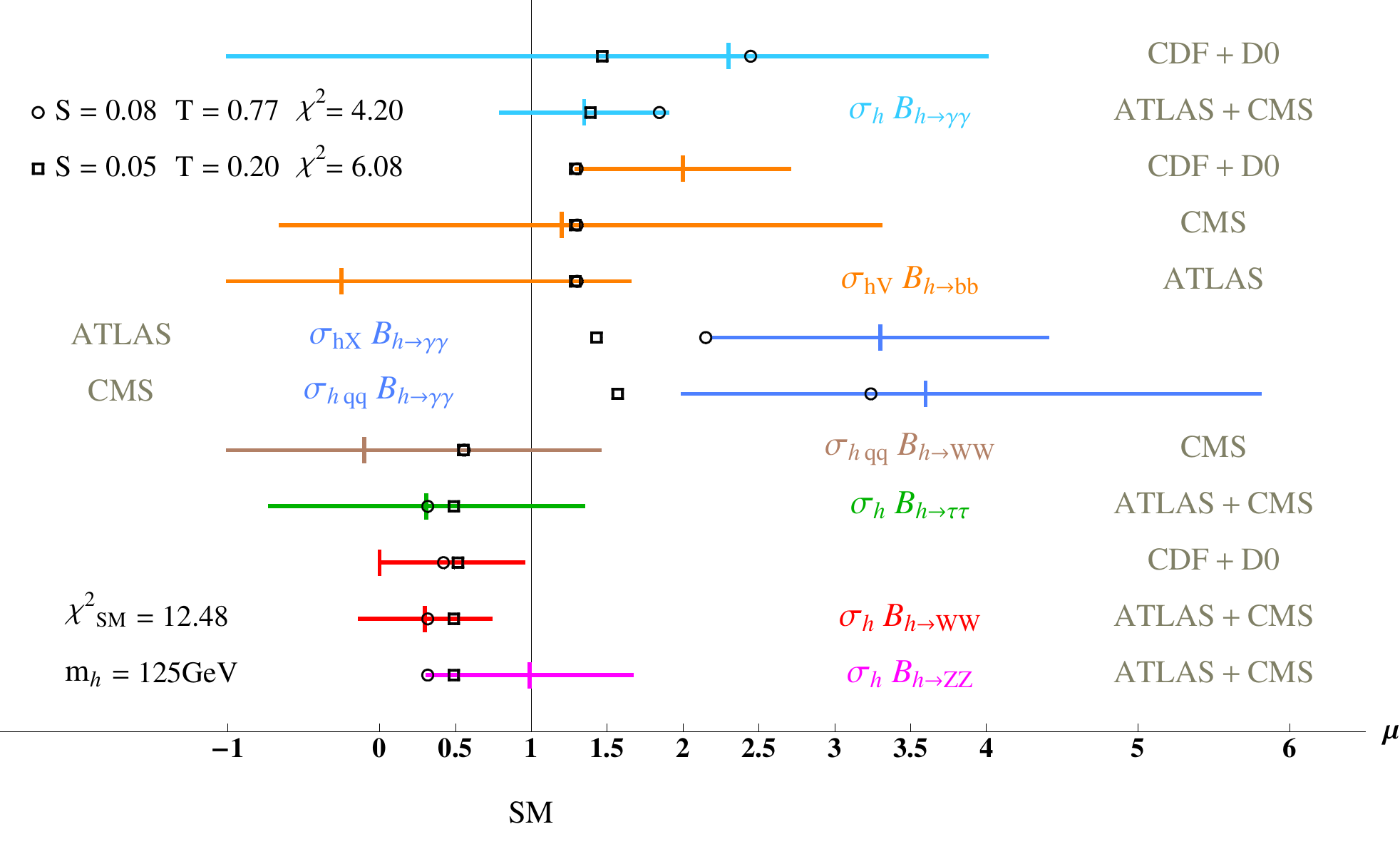}
\\
\end{tabular}
\caption{
\underline{Left:}  
Same as in Fig.(\ref{Fig:Model.II}) but for the Model I with the $(b'',q_{-4/3})$ doublet [see Eq.(\ref{Eq:model.I.2})]   
in the plan $m_{-7/3}$ versus $m'_{-7/3}$ (in GeV). 
The fixed parameters read as, $Y=1$, $Y_{-7/3}=3$, $Y_b =-0.053$, $Y'_b =1$, $Y_{-1/3} =1$, $m'=1200$~GeV, $m_{-4/3}=900$~GeV.
Three contour-level curves for $m_{q^1_{-7/3}}$ are shown.
Other masses are almost constant over the plan, $m_{b_1}\approx 4$~GeV, $m_{b_2}\approx 840$~GeV, $m_{b_3}\approx1290$~GeV,  
$m_{t_1}\approx 173$~GeV, 
and the remaining ones are, $m_{q^2_{-7/3}}\approx 1750 - 2750$~GeV [in the presented domain], 
$m_{q^1_{-4/3}}= min(m_{-4/3},m_{-7/3})$, $m_{q^2_{-4/3}}= max(m_{-4/3},m_{-7/3})$
[as there is no mixing term].
\underline{Right:}  
Same as in Fig.(\ref{Fig:Model.II}) for the experimental data but with theoretical predictions from the Model I [with $(b'',q_{-4/3})$].  
The black squares are associated to the same input parameter values as for the black circles except that 
$Y_{-7/3}=2.2$.}
\label{Fig:Model.I}
\end{center}
\end{figure}

An example of parameter domain leading to Higgs rates within the $1\sigma$ regions is also shown  
for the Model I with the $(b'',q_{-4/3})$ doublet [defined by Eq.(\ref{Eq:model.I.2})] in the left-part of Fig.(\ref{Fig:Model.I}). 
The fundamental parameters are noted in a similar way as for the Model II discussed in previous subsection.
\\ 
In this case, the constraints on $m_{q^1_{-7/3}}$ deserve some more attention. Indeed, the $q^1_{-7/3}$ particle is stable for the reasons exposed in 
Section~\ref{minmodels} that apply here due to the kinematical feature, $m_{q^1_{-7/3}} < m_{-7/3}$ ($m_{-7/3}$ being the mass of the non-mixed
and thus eigenstate $q_{-4/3}$), visible in Fig.(\ref{Fig:Model.I}). 
Now such a stable particle has never been searched at colliders so far, but one can try to extrapolate constraints on it from investigations on other 
Heavy Stable Charged Particles (HSCP) even if this is a non-trivial task; 
constraints have been imposed on long-lived supersymmetric partners -- namely a gluino $\tilde g$, a tau-slepton $\tilde \tau_1$, a top-squark $\tilde t_1$ -- 
and the most stringent ones have been derived recently at the LHC~\cite{Chatrchyan:2012sp} (see Ref.~\cite{Acosta:2002ju} for analog studies at the Tevatron).
After the hadronization stage, a long-lived gluino should form a `R-gluonball' $\tilde g g$ with a probability of typically $\sim 30\%$ (based on simulations) 
and a color-singlet `meson' $\tilde g \bar q q$ ($q\equiv u,d,s$~quarks) with $Q_{e.m.}=\pm 1$ also for a significant fraction, 
whereas a long-lived stop might form a meson $\tilde t_1 \bar q$ ($q=d,s$) with a fraction typically about $\sim 50\%$~\cite{Fairbairn:2006gg}.  
To get an idea of a possible mass bound on the stable $q^1_{-7/3}$ which will also form a $Q_{e.m.}=\pm 1$ R-hadron~\footnote{Inelastic hadronic interactions 
could change the charge-sign of the exotic hadron~\cite{Drees:1990yw}.}, more precisely a baryon $q^1_{-7/3} u u$,
one could apply the mass-dependent limits on the $\tilde g$ ($\tilde t_1$) production rates obtained in Ref.~\cite{Chatrchyan:2012sp} 
to the $q^1_{-7/3}$ pair production cross section calculated at NNLO for the $7$~TeV~LHC. We find, $m_{q^1_{-7/3}} \gtrsim 850$~GeV
($800$~GeV). 
Nevertheless, there are various limitations to the validity of this extrapolation: the fraction of $q^1_{-7/3} u u$ should be simulated specifically
({\it e.g.} with the {\tt PYTHIA} or {\tt HERWIG} simulators) and is certainly different from the $\tilde g \bar q q$ ($\tilde t_1 \bar q$) fraction, while the
spin configuration and/or color-multiplet also differ from each other.  
Furthermore, the nuclear interactions experienced in matter by R-hadrons, suffering from large uncertainties, may lead to charge exchange; 
a recent work~\cite{Mackeprang:2009ad} modeling the HSCP nuclear interactions favors a scenario where the majority of R-hadrons made of a gluino 
or a squark would emerge neutral in the muon detectors.  
Assuming this result extends to all the R-hadrons made of the $q^1_{-7/3}$ ($q^1_{-7/3} q q$, $q^1_{-7/3} \bar q$) and using again the exclusion limits on the gluino (stop) production 
rates~\cite{Chatrchyan:2012sp} obtained now under this charge suppression hypothesis, we find, $m_{q^1_{-7/3}} \gtrsim 750$~GeV ($800$~GeV).  
As for the limit on the $\tilde \tau_1$ production rate, it ends up at a $\sim 500$~GeV~mass~\cite{Chatrchyan:2012sp}. Hence, taking the mean of the above 
four estimated limits, we obtain an indicative bound, $m_{q^1_{-7/3}} \gtrsim 800$~GeV, that is represented over  
the domain drawn in the left-part of Fig.(\ref{Fig:Model.I}); one sees that a large part of the domain is passing this indicative test.
\\ 
There also exist studies on long-lived charged massive particles outside colliders (see {\it e.g.} the reviews in Ref.~\cite{Fairbairn:2006gg} and Ref.~\cite{Perl:2001xi})
leading in particular to limits on their abundance in ordinary matter; these particles can bind to a nucleus forming anomalously (super)heavy isotopes which have been searched, 
or even fall onto/through oceans and lakes to form heavy water molecules. For example, constraints on astrophysical fluxes of those particles have also been looked 
at~\footnote{Let us just mention the following; it has even been proposed some time ago that certain charged massive particles could also 
constitute candidates for the dark matter of the universe (see for instance Ref.~\cite{Goodman:1984dc,Goldberg:1986nk,Dimopoulos:1989hk,DeRujula:1989fe}).}. 
However, no dedicated analysis (outside colliders) 
has really been performed to constrain significantly the mass of stable color-triplet fermions with high fractional electric charges.
\\Ê
The eigenstate $q_{-4/3}$ decays as, $q_{-4/3}\to q^1_{-7/3} W^+$ or $q_{-4/3}\to q^{2(\star)}_{-7/3} W^+ \to q^1_{-7/3} h W^+, q^1_{-7/3} Z^0 W^+$ (open in most of the
considered domain). 
In this case the stable $q^1_{-7/3}$ could also be searched at colliders as missing energy associated with boson  
production, but there were no investigation so far on these kinds of $q_{-4/3}$ decays within the present framework 
[the $q_{-4/3}$ mass, $m_{-7/3}$, has typically high values anyway as shown in Fig.(\ref{Fig:Model.I})]. 
Similarly, the $q^2_{-7/3}$ decays, $q^2_{-7/3}\to q^1_{-7/3} h, q^1_{-7/3} Z^0$ and  
$q^2_{-7/3}\to q^{(\star)}_{-4/3} W^-\to q^1_{-7/3} W^+ W^-$, lead to specific signatures not yet analyzed.
The constraint for the unmixed $q'_{-4/3}$ eigenstate mass, $m_{-4/3}>560$~GeV (since $B_{q'_{-4/3}\to b_1W}\simeq 1$, as possibly for a $t_2$)~\cite{CMS:2012ab}, 
is also clearly respected. Finally, the $b$-sector is exactly as in the Model II of previous subsection so that the positive conclusions for the $b_{2,3}$ mass bounds are similar.

The square-points in the right-part of Fig.(\ref{Fig:Model.I}) for the considered Model I 
correspond to reasonable $S,T$ values at the border of the $1\sigma$ ellipse of EWPT~\cite{Nakamura:2010zzi}.
Those theoretical predictions improve the fits of most of the Higgs rates compared to the SM, leading to a net improvement of the $\chi^2$ value.
The circle-points even all belong to the $1\sigma$ regions with again a remarkable increase for $\mu_{q\gamma}$; 
however, the associated $S,T$ values are now clearly outside the $1\sigma$ 
ellipse~\cite{Nakamura:2010zzi}. Other types of new physics 
effects could reduce the $T$ parameter down to acceptable values -- the $S$ parameter being already in a realistic range -- 
as discussed in the end of Section~\ref{parspace}.

The Models of type III [see Eq.(\ref{Eq:model.III})] lead to similar allowed domains of parameter space as above, as well as comparable qualities of the Higgs rate fits. 
The main difference with previous models concerns once more the type of bound applying to the mass of the VL quark with the highest absolute electric charge, namely here $q^1_{-7/3}$. 
As explained in Section~\ref{minmodels}, it decays predominantly as, $q^1_{-7/3} \to q_{-4/3}^{1(\star)} W^- \to b_1 W^- W^-$, whose final state mimics that of, 
$b_2 \to t_1 W^- \to b_1 W^+ W^-$. In obtained acceptable parameter ranges, the $m_{q^1_{-7/3}}$ values stand inside the interval $\sim 700 - 1000$~GeV
being clearly above the experimental bound, $m_{q^1_{-7/3}} > 611$~GeV ($B_{q^1_{-7/3}\to b_1 W W} \simeq 1$)~\cite{Chatrchyan:2012yea}.  
In the Models III, there are no stable VL quarks.

Of course, we are not going to present detailed numerical results for the various minimal models of types I, II and III presented in Section~\ref{minmodels}, 
but similar qualitative and quantitative results as the ones presented throughout this paper hold.

\section{Conclusions}
\label{conclu}

We have shown that the presence of VL quarks is sufficient to modify the SM Higgs boson rates such that all those belong to the 
present experimental $1\sigma$ regions.
The minimal field contents and gauge group representations of VL quarks allowing to achieve such improvements have been obtained.  
The key idea is to introduce VL quarks with high-enough electric charges to increase sufficiently the diphoton channel rates. Simultaneously, all the
(in)direct constraints can be satisfied, even if we have pointed out a little tension between optimizing the Higgs rate fit and respecting the EWPT. 
The models obtained predict a rich phenomenology at LHC with several exotic-charge quarks possibly around the TeV. 
For example, the pair production of $q^1_{8/3}$, decaying as $q^1_{8/3} \to t_1 W^+ W^+$, could lead to a spectacular final state with six $W$ gauge bosons
(two more $W$'s than in the $q^1_{5/3}$ pair production~\cite{Contino:2008hi}).

One must mention that obviously the present measurements of the Higgs boson rates have large uncertainties. 
If the next data, in particular from the $8$~TeV~LHC, confirm the existence of a $\sim 125$~GeV Higgs field, two typical situations might arise.
First, the main features of the present rate measurements, like diphoton channels significantly larger than in the SM, might be confirmed pointing towards  
manifestations of a theory beyond the SM. In this case, a simple adjustment of the present fundamental parameters should suffice to optimize the Higgs rate fit. 
Even the conclusion on the best-fit configuration underlined here would remain unchanged with respect to the central values; it is 
moreover true for any pure VL quark model (independently of its field content and gauge multiplets).  
A second possibility is that significant changes appear among the Higgs rate measurements so that the new diphoton rates
get closer to their SM expectations. In such a case of agreement, our work could serve as a guide to constrain the VL quark models: for instance,
such new data would disfavor the presence of ${\cal O}(1)$~TeV VL quarks coupled to the Higgs boson and 
with absolute electric charges at $7/3$ or above.
\\
\\
{\bf Acknowledgements:}
G.M. thanks G.~Servant for useful discussions as well as the organizers of the CERN European School of High-Energy Physics 2012 
where this paper was finalized. 
This work is supported by the ``Institut Universitaire de France''. G.M. also acknowledges support from  
the ANR {\it CPV-LFV-LHC} under project \textbf{NT09-508531} and ANR {\it TAPDMS} under project \textbf{09-JCJC-0146}.

\bibliography{VLfitH.4}

\end{document}